\newcommand{\be}[0]{\begin{equation}}
\newcommand{\ee}[0]{\end{equation}}
\newcommand{\ba}[0]{\begin{eqnarray}}
\newcommand{\ea}[0]{\end{eqnarray}}
\documentclass[twoside]{article}
\usepackage{fleqn,espcrc2}
\usepackage{graphicx}

\newcommand{\AmS}{{\protect\the\textfont2
  A\kern-.1667em\lower.5ex\hbox{M}\kern-.125emS}}

\title{EMC effect and nuclear structure functions}

\author{
       {S. Atashbar Tehrani\address[MCSD]{
Physics Department, Persian Gulf University, Boushehr, Iran \\
and Institute for Studies in Theoretical Physics and Mathematics
 , P.O.Box 19395-5531, Tehran, Iran }%
\thanks{e-mail: Atashbar@ipm.ir}, A. Mirjalili\address[MCSD]{
Physics Department, Yazd University, Yazd, Iran \\
and Institute for Studies in Theoretical Physics and Mathematics
 , P.O.Box 19395-5531, Tehran, Iran }%
\thanks{e-mail: Mirjalili@ipm.ir}\\ }and
       {Ali N. Khorramian\address[MCSD]{
Physics Department, Semnan University, Semnan, Iran \\
and Institute for Studies in Theoretical Physics and Mathematics
 , P.O.Box 19395-5531, Tehran, Iran}%
\thanks{e-mail: Alinaghi.Khorramian@cern.ch }}}

\begin{document}

\begin{abstract}
We analyze experimental data of nuclear structure function ratios
$F_2^A/F_2^D$ for obtaining optimum parton distribution functions
(PDFs) in nuclei. Then, uncertainties of the nuclear PDFs are
estimated by the Hessian method. Parametrization of nuclear parton
distribution is investigated in the leading order of $\alpha_s$.
The parton distribution are provided at $Q^2=1 GeV^2$ with a
number of parameters, which are determined by a $\chi^2$ analysis
of the data on nuclear structure function. From the analysis, we
propose parton distributions at $Q^2=1 GeV^2$ for nuclei from
deuteron to heavy ones with a mass number $A\sim 208$.
\end{abstract}

\maketitle

\section{Introduction}

The structure function $F_2$ for a bound nucleon, measured in deep
inelastic scattering (DIS) of leptons, differs from that for a
free nucleon. Unpolarized parton distribution in the nucleon are
now well determined in the region from very small $x$ to large $x$
by using various experimental data. There are abundant data on
electron and muon deep inelastic scattering. In addition, there
are available data from neutrino reactions, Drell-Yan processes,
$W$ production, direct photon production, and others.\\\\
In this paper we determine the unbounded parton distributions and
structure function for free nucleon. For this peropus, we use the
Bernstein average of moments, as we used in
\cite{JHEP:2004,AIP:2004,CTP:2005,khorramIJMP:2005}. The
polynomials are chosen so that the range of $x$ for which the
experimental values of $F_{2}^p$ are not determined, make only a
small contribution to the averages. These experimental Bernstein
averages are then fitted in sec. 2, using CERN subroutine MINUTE
\cite{minuit:3}, to the QCD predictions for the corresponding
linear combinations of moments. By having free parton
distributions we can obtain the bounded parton distributions to
determine nuclear structure function. In section 3 we calculate
the weight function which takes into account the nuclear
modification. Finally our nuclear parametrization studies are
summarized  in last section.
\section{Unbounded parton distribution in $n$ and $x$ space}
The $Q^2$ evolutions of parton distribution functions in moment
space are given by \cite{GGR:1989}\[
 M_{u_v}(n,Q^2)=e^{-d_{NS}s}\times M_{u_v}(n,Q_0^2)\;,\;\;\;\;\;\;\;\;\;\;\;\;\;\;
 \;\;\;\;\;\;\;\;\;\;\;\;\;\;
 \;\;\;\;\;\;\;\;\;\;\;\;\;\;\]
 \[M_{d_v}(n,Q^2)=e^{-d_{NS}s}\times M_{d_v}(n,Q_0^2)\;,\;\;\;\;\;\;\;\;\;\;\;\;\;\;
 \;\;\;\;\;\;\;\;\;\;\;\;\;\;
 \;\;\;\;\;\;\;\;\;\;\;\;\;\;\]
 \ba \Sigma(n,Q^2)=(\alpha_n\Sigma(n,Q_0^2)+\beta_n g(n,Q_0^2))e^{-d_{-}s} \nonumber\;\;\;\;\;\;\;\;\;\;\;\;\;\;
 \;\;\;\;\;\;\;\;\;\;\;\;\;\;\\
 +((1-\alpha_n)\Sigma(n,Q_0^2)-\beta_n g(n,Q_0^2))e^{-d_{+}s}
 \nonumber\;,
 \;\;\;\;\;\;\;\;\;\;\;\;\;\;\;\;\;\;\;
 \;\;\;\;\;\;
\ea
 \ba g(n,Q^2)=((1-\alpha_n)g(n,Q_0^2)\;\;\;\;\;\;\;\;\;\;\;\;\;\;\;\;\;\;\;\;\;\;\;\;\;\;\;\;\;\;\;\;\;\;\;\;\;\;\;\;\;\;
 \;\;\;\;\;\;\;\;\;\;
 \nonumber\\+\frac{\alpha_n(1-\alpha_n)}{\beta_n}\Sigma(n,Q_0^2))e^{-d_{-}s}
\;\;\;\;\;\;\;\;\;\;\;\;\;\;
 \;\;\;\;\;\;\;\;\;\;\;\;\;\;\;\;\;\;\;\;\;\;\;\;
 \;\;\;\;\;
 \nonumber\\
 +(\alpha_n g(n,Q_0^2)-\frac{\alpha_n(1-\alpha_n)}{\beta_n}\Sigma(n,Q_0^2))e^{-d_{+}s}
    \;. \;\;\;\;\;\;\;\;\;\;\;\;\;\;\;\;\;\;\;\; \nonumber \ea
     \ba \label{eq:moment}\ea

Here $\alpha_n$ and $\beta_n$ and associate anomalous dimension
are as following
\ba \alpha_n&=&\frac{d_{ns}-d_{+}}{d_{-}-d{+}}\;,\nonumber\\
\beta_n&=&\frac{d_{qg}}{d_{-}-d{+}}\;,\nonumber\\
d_{+}&=&\frac{1}{2}(d_{NS}+d_{gg}+\Delta)\;,\nonumber\\
d_{-}&=&\frac{1}{2}(d_{NS}+d_{gg}-\Delta)\;,\nonumber\\
\Delta&=&\sqrt{(d_{NS}-d_{gg})^2+4d_{gq}d_{qg}}\;,\nonumber\\
d_{NS}&=&\frac{1}{3\pi b}\left(1-\frac
{2}{n(n+1)}+4\sum_{j=2}^n\frac{1}{j}\right)\;,\nonumber\\
d_{gq}&=&\frac{-2}{3\pi b}\frac{2+n+n^2}{n(n^2-1)}\;,\nonumber\\
d_{qg}&=&\frac{-f}{2\pi b}\frac{2+n+n^2}{n(n+1)(n+2)}\;,\nonumber\\
d_{gg}&=&\frac{-3}{\pi b}(-\frac{1}{12}+\frac{1}{n(n-1)}+\frac{1}{(n+1)(n+2)}\;,\nonumber\\
&-&\frac{f}{18}-\sum_{j=2}^n\frac{1}{j})\;,\nonumber\\
\label{eq:alamos} \ea the parameter $s$ and $b$ in above equations
are defined as
 \ba s&=&ln\frac{ln\frac{Q^2}{\Lambda^2}}{ln\frac{Q_0^2}{\Lambda^2}}\;,\nonumber\\
b&=&\frac{33-2f}{12\pi}\label{eq:sandb}\;.\ea If we back to
Eq.~(\ref{eq:moment}), we need to know all of the moments of
parton at $Q^2=Q_0^2$.\\\\
 In  phenomenological investigations of structure functions,
for a given value of $Q^2$, only  a limited number of experimental
points,
 covering a partial range of values of $x$, are available. Therefore,
one cannot directly determine the moments.
 A method devised to deal with this situation
 is to take  averages of the structure function weighted by suitable polynomials.
 We can compare theoretical predictions with experimental results for the Bernstein
 averages.
 After extracting some unknown parameters which exist in  the functional form  of input
 parton distribution in moment space, we are able to determine parton distributions in
 $x$-space.
The parameterizations of the parton densities at input scale of
$Q_0^2=1\; GeV^2$ are determined as \ba xu_v(x,Q_0^2)&=&0.462
x^{0.291}(1-x)^{3.219}\nonumber\\&&(1-0.959\sqrt{x}+19.423x)
\;,\nonumber\\
xd_v(x,Q_0^2)&=&0.655 x^{0.424}(1-x)^{4.093}\nonumber\\&&(1-0.634\sqrt{x}+7.315x)\;,\nonumber\\
x\Sigma(x,Q_0^2)&=&0.534 x^{-0.176}(1-x)^{3.286}\nonumber\\
&&(1+0.790 \sqrt{x}+8.698x+11.530x^2)\;,\nonumber\\
xg(x,Q_0^2)&=&2.113 x^{-0.030}(1-x)^{6.403}\nonumber\\
&&(1-2.639\sqrt{x}+9.358x)\;.\ea
 We can control some parameter at input scale in above equations by these
 constrains
 \ba
 &&\int^1_0u_v(x,Q^2_0)dx=2\;,\nonumber\\
 &&\int^1_0d_v(x,Q^2_0)dx=1\;,\nonumber\\
&&\int^1_0(x\Sigma(x,Q^2_0)+xg(x,Q^2_0))dx=1\;.
  \ea
By taking the  moments of the above equation and inserting them in
the Eq.~(\ref{eq:moment}), we access to  all of the parton
distributions in moment space. Then, by using the inverse Mellin
 technique as
 \ba
 f(x,Q^2)&=&\frac{1}{\pi}\int_0^{5+10/ln\frac{1}{x}}dz Im[e^{i\varphi}x^{-c-ze^{i\varphi}}\nonumber\\
  &&M(n=c+ze^{i\varphi},Q^2)]\label{eq:invmelin}
 \ea
 we can obtain  the parton distribution in $x$-space for free proton.
 In above analysis, we chose the $\Lambda=0.22\;GeV$ and c=1.1 for
 non singlet part and c=2.1 for singlet part \cite{GRV:90}.

\begin{figure}[tbh]
\centerline{\includegraphics[width=0.5\textwidth]{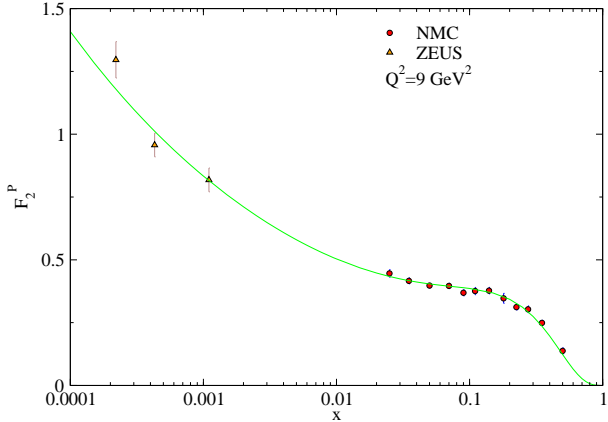}}
\caption{Structure function for free proton in $Q^2=9 GeV^2$ .}
\label{fig:f2p}
\end{figure}

\section{Nuclear parton distributions and structure functions}
Available nuclear data on the structure function $F_2^A$ are taken
in fixed-target experiments at this stage, and they are shown in
Fig. 1 as a function of $Q^2$ and $x=Q^2/(2\;m\;\nu)$, where $\nu$
is the  transferred  energy to the target, $m$ is the nucleon
mass, and $Q^2$ is given by $Q^2\equiv-q^2$ with the virtual
photon momentum $q$. The initial nuclear parton distributions are
provided at a fixed $Q^2(\equiv Q_0^2)$, and they are taken as
\cite{Hirari:0104} \be \label{eq:waightfunction}
f_i^A(x,Q_0^2)=w_i(x,A,z)f_i(x,Q_0^2)\;, \ee where
$f_i^A(x,Q_0^2)$
 is the parton distribution with type $i$ in
the nucleus $A$ and $f_i(x,Q_0^2)$ is the corresponding parton
distribution in the nucleon. We call $w_i(x,A,z)$ a weight
function, which takes into account the nuclear modification. This
functional form should be considered by: \ba
w_i(x,A,Z)&=&1+\frac{\left(1-\frac
{1}{A^{1/3}}\right)}{(1-x)^{\beta_i}}(\alpha_i+a_i\sqrt{x}\nonumber\\
&+&b_ix+c_ix^2+d_ix^3+e_ix^4)\label{eq:waight}\;. \ea Because the
valence-quark distributions in a nucleus are much different from
the ones in the proton, we should be careful in defining the
weight function $w_i(x,A,Z)$.\\ \\
For nuclear $A$ at $Q_0^2$ we have
\[
u_v^A(x,Q_0^2)=w_u(x,A,Z)\frac{Zu_v(x,Q_0^2)+Nd_v(x,Q_0^2)}{A}\;,\]\[
d_v^A(x,Q_0^2)=w_d(x,A,Z)\frac{Zd_v(x,Q_0^2)+Nu_v(x,Q_0^2)}{A}
\;,\]

\[
\Sigma^A(x,Q_0^2)=w_{\Sigma}(x,A,Z)\Sigma(x,Q_0^2) \;,\] \be
g^A(x,Q_0^2)=w_{g}(x,A,Z)g(x,Q_0^2)\;,\ee

\section{Summery}
In the theoretical calculations, the nuclei
are assumed as:\\
$^4He$,$^7Li$,$^9Be$,$^{12}C$,$^{14}N$,$^{27}Al$,$^{40}Ca$,$^{56}Fe$,$^{63}Cu$,$^{107}Ag$,
$^{118}Sn$,$^{131}Xe$,$^{197}Au$, and $^{208}Pb$\\
The initial nuclear distributions are provided at $Q_0^2=1\;GeV^2$
with the parameters in Eq.~(\ref{eq:waight}). To  obtain  some
unknown parameters which appeared in Eq.~(\ref{eq:waight}), we can
use the ratios $R_{F_2}^A=F_2^A(x,Q^2)/F_2^D(x,Q^2)$  to calculate
\be\label{eq:chi}
\chi^2=\sum_j\frac{(R_{F_2,j}^{A,data}-R_{F_2,j}^{A,theory})^2}{(\sigma_j^{data})^2}\;.
 \ee
 where the experimental error is given by considering the systematic and
 statistical errors. Information about the used experimental data is given in
Table I, where nuclear species, references, and data numbers are
listed.
 \begin{center}
 \begin{tabular}{cccc}
\hline\hline Nucleus & Experiment & Reference & No. of data
\\ \hline
He & SLAC-E139 & \cite{E139:94} & 21  \\
& NMC-95 & \cite{NMC:95} & 18 \\
C & EMC-88 & \cite{EMC:88} & 9 \\
& EMC-90 & \cite{EMC:90} & 13 \\
& SLAC-E139 & \cite{E139:94} & 17  \\
& NMC-95 & \cite{NMC:95}& 18 \\
& FNAL-E665-95 & \cite{E665:95} & 11  \\
Ca & EMC-90 & \cite{EMC:90} & 13  \\
& NMC-95 & \cite{NMC:95} & 18 \\
& SLAC-E139 & \cite{E139:94}& 17 \\
& FNAL\_E665-95 & \cite{E665:95} & 11 \\ \hline
\end{tabular}
\end{center}
\textbf{Table. I} Nuclear species`references, and data numbers are
listed for the used experimental data with $Q^2\gg 1\;GeV^2$.\\

 With these preparations together with the CERN subroutine MINUIT \cite{minuit:3}, the
 optimum parameter set is obtained by minimizing $\chi^2$. The experimental data
 are taken from the publications by the European Muon
 Collaboration (EMC)\cite{EMC:88} \cite{EMC:90} at the European
 Organization for Nuclear Research (CERN), E139 Collaborations \cite{E139:94} at
 the Stanford Linear Accelerator Center (SLAC), the New Muon
 Collaboration (NMC) \cite{NMC:95} at CERN, and the E665
 Collaboration \cite{E665:95} at the Fermi National Accelerator
 Laboratory (FNAL). The used data  are for the following nuclei:
 helium (He),carbon (C) and calcium (Ca). In Table II. we present
 the $\chi ^{2}/d.o.f$ for helium, carbon and calcium
 as an example.

\begin{center}
\begin{tabular}{ccc}
\hline\hline Nucleus & No.of data & $\chi ^{2}/d.o.f$ \\ \hline
He & 39 & 1.12 \\
C & 68 & 1.49 \\
Ca & 59 & 3.86 \\ \hline
\end{tabular}
\end{center}
\textbf{Table. II}  The  $\chi^2$ values, resulted from fitting,  for different types of nucleus .\\\\

\begin{figure}[tbh]
\centerline{\includegraphics[width=0.5\textwidth]{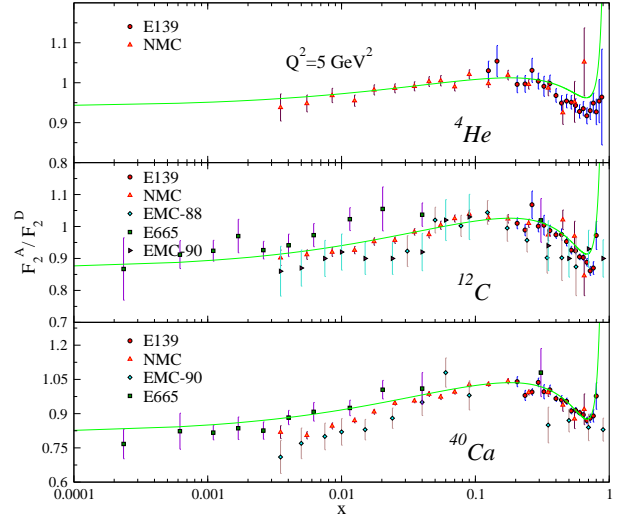}}
\caption{EMC effect for He, C, and Ca nucleus .} \label{fig:emc}
\end{figure}

 In Fig.~(\ref{fig:emc}) we present
 $F_2^{He}(x,Q^2)/F_2^D(x,Q^2)$, $F_2^C(x,Q^2)/F_2^D(x,Q^2)$ and $F_2^{Ca}(x,Q^2)/F_2^D(x,Q^2)$
 as a function of $x$ and for $Q^2=5\; GeV^2$. As it can be seen , these ratios are in good
 agreement to the available experimental data.

\section{Acknowledgments}
S. A. T. acknowledge the Persian Gulf  university for financially
supporting this project.We are grateful to the  Institute for
 studies in theoretical Physics and Mathematics (IPM)  for its
hospitality whilst this research  was performed.

\end{document}